\documentstyle[psfig,amssymb]{article}

\title{Mass Segregation in Young LMC Clusters}

\author{Richard de Grijs\footnote{Present address: Department of Physics
\& Astronomy, University of Sheffield, Hicks Building, Hounsfield Road,
Sheffield S3 7RH, UK}, Gerry F.  Gilmore,\\
{\it Institute of Astronomy, University of Cambridge, Madingley Road,}\\
{\it Cambridge CB3 0HA, UK}\\ ~\\
Rachel A.  Johnson \\
{\it European Southern Observatory, Casilla 19001, Santiago 19, Chile}}

\date{\ }

\begin{document}
\maketitle

\begin{abstract} 
We present the detailed analysis of {\sl Hubble Space Telescope}
observations of the spatial distributions of different stellar species
in two young compact star clusters in the Large Magellanic Cloud (LMC),
NGC 1805 and NGC 1818.  Based on a comparison of the characteristic
relaxation times in their cores and at their half-mass radii with the
observed degree of mass segregation, it is most likely that significant
primordial mass segregation was present in both clusters, particularly
in NGC 1805.  Both clusters were likely formed with very similar initial
mass functions (IMFs).  In fact, we provide strong support for the
universality of the IMF in LMC clusters for stellar masses $m_* \gtrsim
0.8 M_\odot$. 
\end{abstract}

\section{Strong mass segregation on short time-scales}

One of the major uncertainties in modern astrophysics is the issue of
whether the stellar initial mass function (IMF) is universal or,
alternatively, determined by environmental effects.  Galactic globular
clusters and rich, compact Magellanic Cloud star clusters are ideal
laboratories for providing strong constraints on the universality of the
IMF, in particular because they are essentially single age, single
metallicity systems for which statistically significant samples of
individual stars over a range of masses can easily be resolved. 

Although the standard picture, in which stars in dense clusters evolve
rapidly towards a state of energy equipartition through stellar
encounters -- with the corresponding mass segregation -- is generally
accepted, observations of various degrees of mass segregation in very
young star clusters (e.g., de Grijs et al.  2002a,b and references
therein) suggest that at least some of this effect is related to the
process of star and star cluster formation itself.  

The effects of mass segregation in star clusters, with the more massive
stars being more centrally concentrated than the lower-mass stars,
clearly complicates the interpretation of an observed luminosity
function (LF) at a given position within a star cluster in terms of its
IMF.  Without reliable corrections for the effects of mass segregation,
hence for the structure and dynamical evolution of the cluster, it is
impossible to obtain a realistic global cluster LF.  Quantifying the
degree of actual mass segregation is thus crucial for the interpretation
of observational luminosity and mass functions (MFs) in terms of the
IMF, even for very young star clusters. 

The time-scale for the onset of significant dynamical mass segregation 
is comparable to the cluster's dynamical relaxation time (Spitzer \&
Shull 1975, Inagaki \& Saslaw 1985, Bonnell \& Davies 1998, Elson et al.
1998).  A cluster's characteristic time-scale may be taken to be its
half-mass (or median) relaxation time, i.e., the relaxation time at the
mean density for the inner half of the cluster mass for cluster stars   
with stellar velocity dispersions characteristic for the cluster as a   
whole (Spitzer \& Hart 1971, Lightman \& Shapiro 1978, Meylan 1987,
Malumuth \& Heap 1994, Brandl et al.  1996), and can be written as
(Meylan 1987):
\begin{equation}
\label{trelax.eq}
t_{\rm r,h} = {(8.92 \times 10^5)} {M_{\rm tot}^{1/2} \over {\langle m
\rangle}} {R_{\rm h}^{3/2} \over {\log (0.4 \; M_{\rm tot} / \langle m
\rangle})} {\rm yr,}
\end{equation}
where $R_{\rm h}$ is the half-mass (median) radius (in pc), $M_{\rm
tot}$ the total cluster mass, and $\langle m \rangle$ the typical mass 
of a cluster star (both masses in $M_\odot$).

Although the half-mass relaxation time characterises the dynamical      
evolution of a cluster as a whole, significant differences are expected
locally within the cluster.  From Eq.  (\ref{trelax.eq}) it follows
immediately that the relaxation time-scale will be shorter for
higher-mass stars (greater $\langle m \rangle$) than for their
lower-mass companions.  From this argument it follows that
dynamical mass segregation will also be most rapid where the local
relaxation time is shortest, i.e., near the cluster centre (cf.  Fischer
et al.  1998, Hillenbrand \& Hartmann 1998).  The relaxation time in the
core can be written as (Meylan 1987):
\begin{equation}
\label{tcore.eq}
t_{\rm r,0} = {(1.55 \times 10^7)} {{v_s R_{\rm core}^2} \over {\langle
m_0 \rangle \log (0.5 \; M_{\rm tot} / \langle m \rangle})} {\rm yr,}
\end{equation}
where $R_{\rm core}$ is the cluster core radius (in pc), $v_s$ (km
s$^{-1}$) the velocity scale, and $\langle m_0 \rangle$ the mean mass
(in $M_\odot$) of all particles in thermal equilibrium.

It should be kept in mind, however, that even the concept of a ``local
relaxation time'' is only a general approximation, as dynamical
evolution is a continuing process.  The time-scale for a cluster to lose
all traces of its initial conditions also depends on the smoothness of
its gravitational potential, i.e.  the number of stars (Bonnell \&
Davies 1998), the degree of equipartition reached (e.g., Hunter et al. 
1995), and the slope of the MF (e.g., Lightman \& Shapiro 1978, Inagaki
\& Saslaw 1985, Pryor et al. 1986, Sosin 1997), among others. 

In addition, as the more massive stars move inwards towards the cluster
centre, their dynamical evolution will speed up.  This process will be
accelerated if there is no (full) equipartition (cf.  Inagaki \& Saslaw
1985), thus producing high-density cores very rapidly, where stellar
encounters occur very frequently and binary formation is thought to be
very effective (cf.  Inagaki \& Saslaw 1985, Elson et al.  1987).  This
may accelerate the mass segregation even more significantly (e.g., Nemec
\& Harris 1987, De Marchi \& Paresce 1996, Bonnell \& Davies 1998, Elson
et al.  1998).  This process will act on similar (or slightly shorter)
time-scales as the conventional dynamical mass segregation (cf.  Nemec
\& Harris 1987, Bonnell \& Davies 1998, Elson et al.  1998). 

\section{Mass segregation and implications for the IMF}

We obtained F555W and F814W {\sl HST/WFPC2} imaging
observations of two young compact LMC clusters, NGC 1805 ($\sim 10$ Myr)
and NGC 1818 ($\sim 25$ Myr), covering a large range of radii (see de
Grijs et al.  2002a for observational details).  The radial dependence
of the LF and MF slopes indicate clear mass segregation in both clusters
at radii $r \lesssim 3-6 R_{\rm core}$ (de Grijs et al.  2002a,b). 

In Fig.  1a we show the dependence of the cluster core radius on the
adopted magnitude (mass) range.  For both clusters we clearly detect the
effects of mass segregation for stars with masses $\log (m/M_\odot)
\gtrsim 0.2$ ($m \gtrsim 1.6 M_\odot$).  Stars with masses $\log
(m/M_\odot) \gtrsim 0.4 \, (m \gtrsim 2.5 M_\odot)$ show a similar
concentration, while a trend of increasing core radius with decreasing
mass (increasing magnitude) is apparent for lower masses. 

Elson et al.  (1987) estimated the central velocity dispersion in NGC
1818 to be in the range $1.1 \lesssim \sigma_0 \lesssim 6.8$ km
s$^{-1}$.  Combining this central velocity dispersion, the core radius
of $\simeq 2.6$ pc (de Grijs et al.  2002a,b), and the cluster age of
$\simeq 25$ Myr, we estimate that the cluster core is between $\sim 5$
and $\sim 30$ crossing times old, so that dynamical mass segregation in
the core should be well under way.  Although we do not have velocity
dispersion information for NGC 1805, it is particularly interesting to
extend this analysis to this younger ($\sim 10$ Myr) cluster.  We know
that its core radius is roughly half that of NGC 1818 (de Grijs et al. 
2002a,b), and its mass is a factor of $\sim 10$ smaller.  Simple scaling
of Eq.  (\ref{trelax.eq}) shows then that the half-mass relaxation time
of NGC 1805 is $\sim 4 - 5\times$ as short as that of NGC 1818; if we
substitute the scaling laws into Eq.  (\ref{tcore.eq}), we estimate that
the central velocity dispersion in NGC 1805 is $\gtrsim 10\times$
smaller than that in NGC 1818.  From this argument it follows that the
cluster core of NGC 1805 is $\lesssim 3-4$ crossing times old. 

However, since strong mass segregation is observed out to $\sim 6 R_{\rm
core}$ and $\sim 3 R_{\rm core}$ in NGC 1805 and NGC 1818, respectively,
for stellar masses in excess of $\sim 2.5 M_\odot$, it is most likely
that significant primordial mass segregation was present in both
clusters, particularly in NGC 1805 (cf.  Fig.  1b). 

\begin{figure}
\parbox{6cm}{\psfig{figure=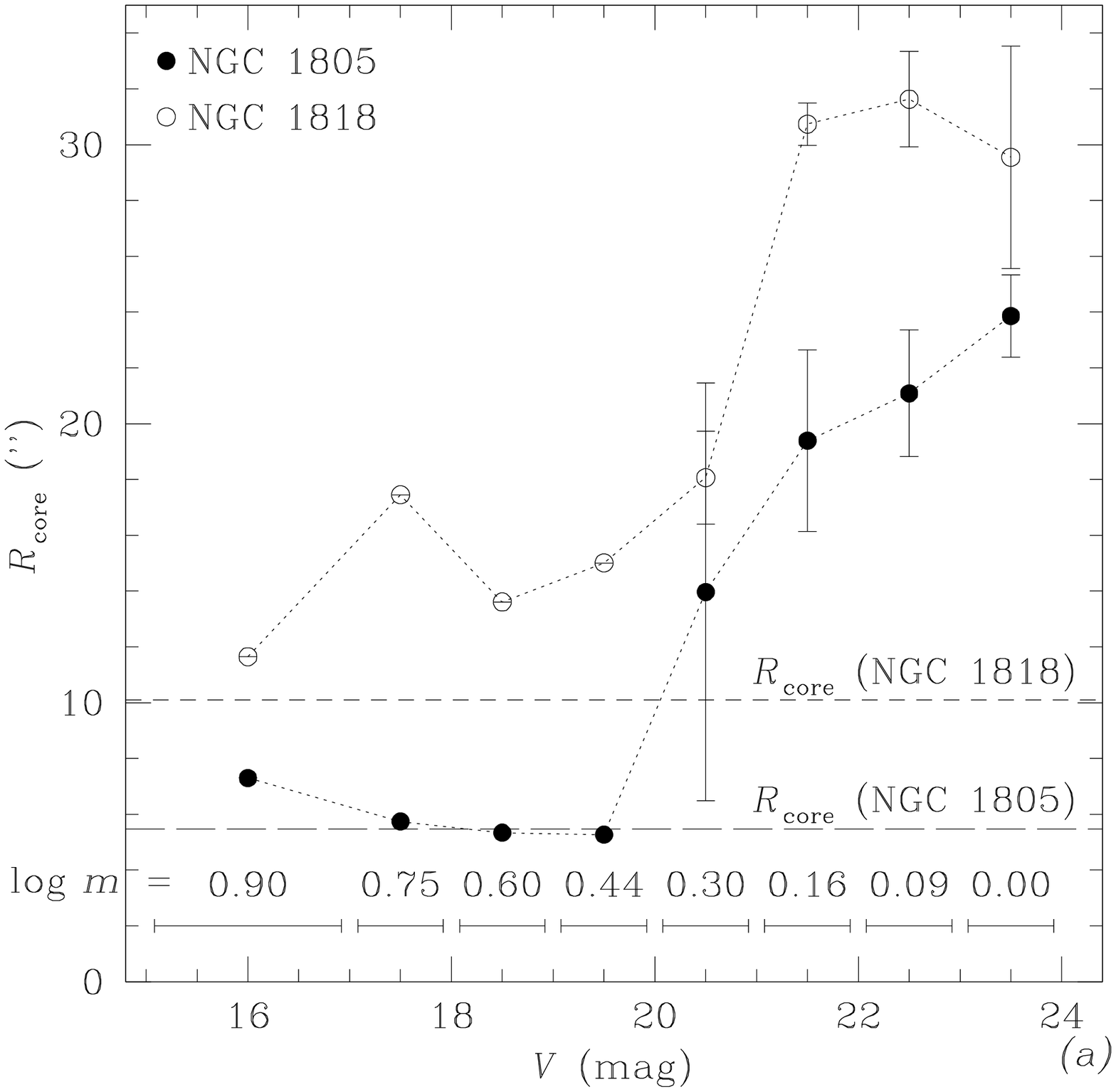,width=6cm}}
\parbox{6cm}{\psfig{figure=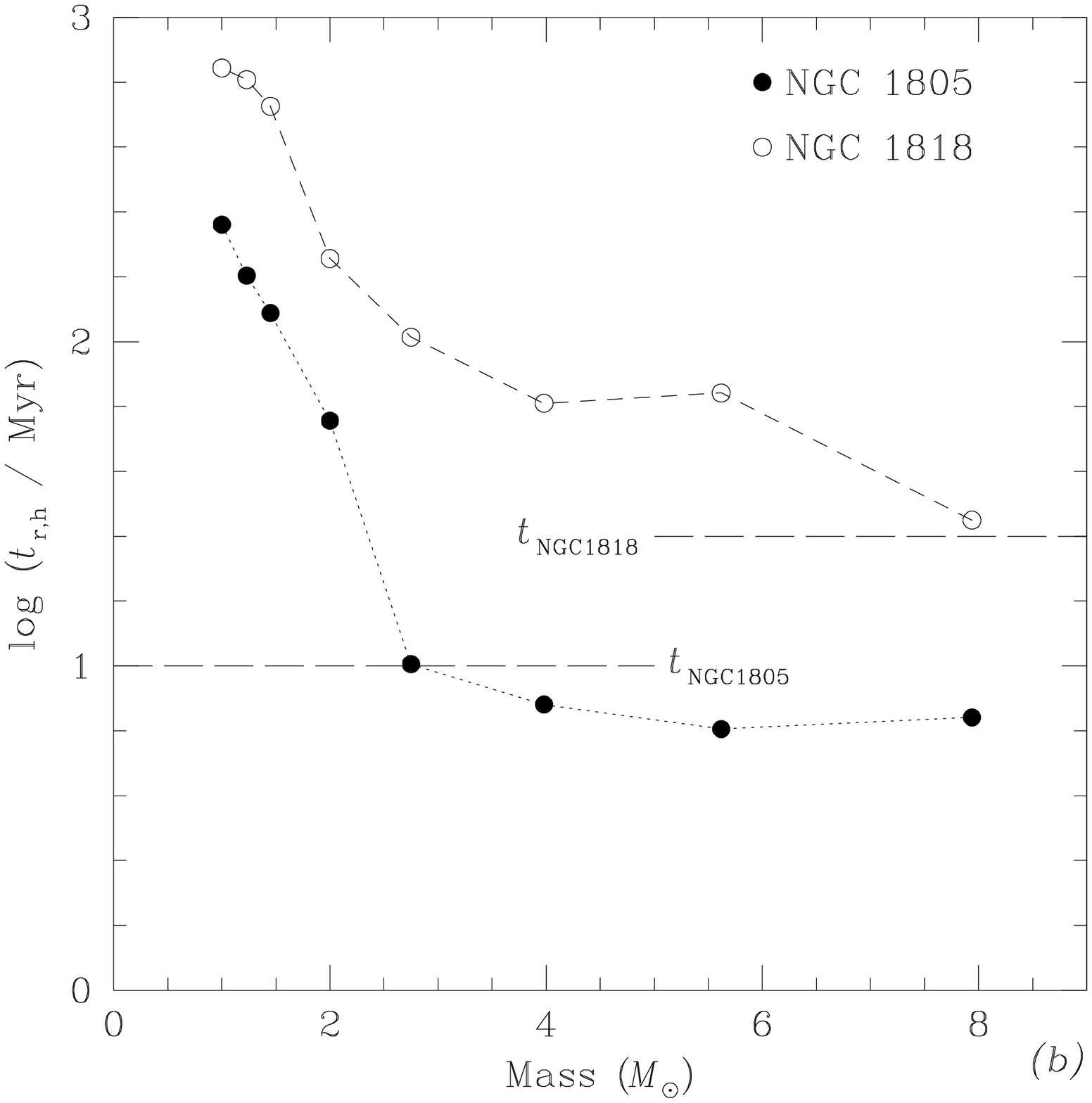,width=6cm}}
\caption{{\it (a)} -- Core radii as a function of magnitude (mass) for
both clusters.  The error bars are driven by uncertainties in the
background subtraction; fitting ranges are indicated at the bottom of
the panel.  The horizontal dashed lines represent the overall core radii
from the clusters' surface brightness profiles.  {\it (b)} -- Half-mass
relaxation time as a function of mass for NGC 1805 and NGC 1818.  The
best age estimates for both clusters are indicated by horizontal dashed
lines.}
\end{figure}

Within the uncertainties, we cannot claim that the slopes of the outer
MFs in NGC 1805 and NGC 1818 are significantly different, which
therefore implies that these clusters must have had very similar IMFs. 
In fact, in de Grijs et al.  (2002c) we extended our study of mass
segregation in clusters of various ages to a sample of six rich LMC
clusters, selected to include three pairs of clusters of similar age
(roughly $10^7, 10^8$ and $10^9$ yr old), metallicity, and distance from
the LMC centre, and exhibiting a large spread in core radii between the
clusters in each pair.  The large spread in core radii in any given
cluster pair was chosen because the core radius distribution of rich LMC
clusters systematically increases in both upper limit and spread with
increasing cluster age (e.g., Mackey \& Gilmore 2002 and references
therein).

All clusters show clear evidence of mass segregation: (i) their
luminosity function slopes steepen with increasing cluster radius, and
(ii) the brighter stars are characterized by smaller core radii.  {\it
For all sample clusters}, both the slope of the luminosity function in
the cluster centres and the degree of mass segregation are similar to
each other, within observational errors of a few tenths of power-law
slope fits to the data.  This implies that their {\it initial} mass
functions must have been very similar, down to $\sim 0.8 - 1.0 M_\odot$
(cf.  de Grijs et al.  2002c). 

We therefore rule out IMF variations as the main driver of the
increasing spread of cluster core radii with age (e.g., Elson et
al.  1989).

\end{document}